# First hydrogen operation of NIO1: characterization of the source plasma by means of an optical emission spectroscopy diagnostic [a]

M. Barbisan,[1,b] C. Baltador,[1] B. Zaniol,[1] M. Cavenago,[2] U. Fantz,[3] R. Pasqualotto,[1] G. Serianni[1], L.Vialetto[4] and D. Wünderlich[3]

[1] Consorzio RFX (CNR, ENEA, INFN, Univ. of Padua, Acciaierie Venete SpA), C.so Stati Uniti 4 – 35127, Padova (Italy)
[2] INFN-LNL, v.le dell'Università 2, I-35020, Legnaro (Italy)
[3] Max-Planck Institut für Plasmaphysik, Boltzmannstr. 2, 85748 Garching, Germany
[4] Università degli Studi di Padova, Via 8 Febbraio, 2 - 35122 Padova

NIO1 is a compact and flexible radiofrequency $H^-$ ion source, developed by Consorzio RFX and INFN-LNL. Aim of the experimentation on NIO1 is the optimization of both the production of negative ions and their extraction and beam optics. In the initial phase of its commissioning, NIO1 was operated with nitrogen, but now the source is regularly operated also with hydrogen. To evaluate the source performances an optical emission spectroscopy diagnostic was installed. The system includes a low resolution spectrometer in the spectral range of 300-850 nm and a high resolution (50 pm) one, to study respectively the atomic and the molecular emissions in the visible range. The spectroscopic data have been interpreted also by means of a collisional-radiative model developed at IPP Garching. Besides the diagnostic hardware and the data analysis methods, the paper presents the first plasma measurements across a transition to the full H mode, in a hydrogen discharge. The characteristic signatures of this transition in the plasma parameters are described, in particular the sudden increase of the light emitted from the plasma above a certain power threshold.
[DOI]

## I. INTRODUCTION

Neutral Beam Injectors (NBIs) will have a relevant role for the heating and current drive of the ITER experiment and the DEMO reactor [1]. The state-of-the-art concept [2] for the production of high energy neutrals foresees the generation of $H^-/D^-$ ions by a cesiated RF source; the negative ions are electrostatically accelerated and then neutralized by charge exchange with $H_2/D_2$ molecules. The production process of the neutral beam still has large margins of optimization, in terms of the final beam quality (uniformity, divergence, etc.) but also of the overall energetic efficiency of NBIs. The NIO1 experiment is devoted to optimize the production, acceleration and neutralization of negative ions, with the development and test of new concepts for the future DEMO NBI. NIO1 is composed by a RF negative ion source, coupled to a 3 grid acceleration system. The extracted beam will be made by 9 beamlets of $H^-$ ions accelerated up to 60 keV, for a total current of maximum 130 mA [3]. At present, the plasma operations of the source have been tested using both air and hydrogen, but with no beam extraction. An RF input power of 1.7 kW (2.5 kW is the maximum for the installed RF generator) has been successfully coupled to the plasma, and the source filling pressure has been reduced down to 1 Pa. The present work illustrates the preliminary results obtained by Optical Emission Spectroscopy (OES) diagnostics during the first NIO1 hydrogen campaign. The plasma properties were studied by analyzing the spectrum of the plasma light in the range 300-850 nm, by means of a low and a high resolution spectrometer, receiving light from Lines Of Sight (LOSs) parallel and close to the Plasma Grid (PG). The interpretation of the experimental data was performed with the help of YACORA, a Collisional-Radiative (CR) model developed at IPP Garching capable to simulate the population coefficients of H and $H_2$ excited states for low pressure and low temperature hydrogen plasmas [4]. It has been possible to measure the rotational temperature of the $H_2$ molecules, the electron temperature and density, the dissociation degree of the molecules and the ionization degree of the atoms. OES has been routinely applied on NIO1, being the sole diagnostic capable of measuring the plasma response to the different operative conditions set by the operator. After a brief description of NIO1 OES diagnostic, the experimental results will be presented and discussed.

## II. DIAGNOSTIC SETUP

In NIO1, the light emitted from the plasma has been observed from two viewports at 26 mm from the PG; they look exactly one into each other, so that the LOSs have collected light from the same region. Each of the 2 viewports hosts an optic head, consisting of a 50 mm focal

---

length BK7 lens, which conveys the plasma light into a quartz optical fiber. The clear aperture of the lens is 4 mm to avoid the vignetting from the viewports' pipes, whose internal diameter is 8 mm. The optic heads have been focused in order to obtain a well collimated LOS along the entire plasma [5]. One optic head is connected through the optical fiber to a low resolution spectrometer *Hamamatsu C10082CAH* [6], mounting an integrated back thinned CCD sensor of 2048 pixels. The device has a plate factor of 0.30÷0.36 nm/pix and a resolution of 1 nm. This spectrometer is dedicated to the acquisition of Balmer series, but it is also helpful to monitor levels of possible impurities in the plasma and the regular operation of NIO1 without noticeable sputtering of source wall materials.. The other optic head is connected to a high resolution spectrometer *Acton SpectraPro-750* [7], with a 2D back illuminated frame transfer CCD camera of 512x512 pixel, for a spectral window at 650 nm, 6 nm wide, a plate factor of 0.012 nm/pix and a resolution of 50 pm (entrance slit set to 70 µm). Both the spectrometers have been absolutely calibrated. The usable spectral window of the diagnostic is from 300 nm to 850 nm, with the lower limit set by the low transmissivity of lenses and fibers in the UV range.

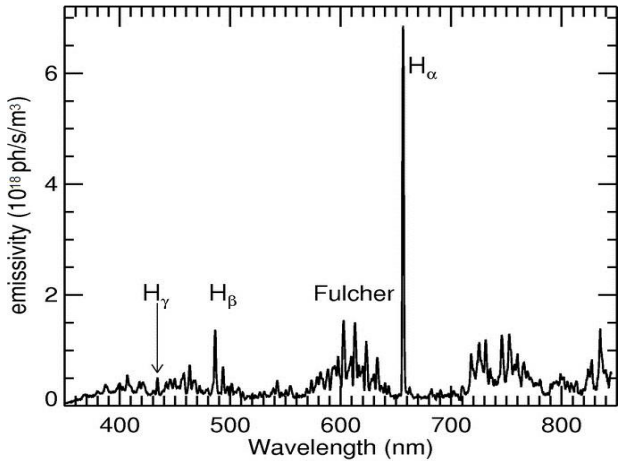

**Figure 1:** Spectrum recorded by the low resolution spectrometer, in emissivity units. The source had an internal pressure of 2.4 Pa and was fed by 1575 W RF power. The exposure time of the spectrometer CCD was of 1 s.

### III. EXPERIMENTAL RESULTS AND ANALYSIS

A typical spectrum acquired by the low resolution system is shown in fig.1. The spectrum refers to plasma produced in NIO1 source with an input RF power of 1575 W and a gas pressure of 2.4 Pa. The most intense Balmer lines ($H_\alpha$, $H_\beta$ and $H_\gamma$) are clearly visible in the spectrum, as the molecular $H_2$ emission of the Fulcher band, mostly comprised in the wavelength range from 595 to 640 nm. The Fulcher band has been acquired also by the high resolution system to distinguish the lines composing it. These have been analyzed to measure the rotational temperature $T_{rot}$ of the $H_2$ molecules, which can be considered a good estimation of the gas temperature [8].

$T_{rot}$ has been calculated from the intensity of the Q1-3 lines of the $d^3\Pi_u \rightarrow a^3\Sigma_g^+$ transition following [8].

The measurements of electron temperature $T_e$, electron density $n_e$, dissociation degree $n_H/n_{H2}$ ($n_H$ and $n_{H2}$ are the densities of atomic and molecular hydrogen) and ionization degree $\alpha$ have been obtained with a recursive algorithm over $T_e$. Firstly, the electron density was calculated from comparing the experimental ratio of $H_\beta$ and $H_\gamma$ emissivities, $\varepsilon_{H\beta}$, $\varepsilon_{H\gamma}$, exploiting the predictions provided by the YACORA model [4], which depend on electron temperature and density. An initial electron temperature of 2 eV has been set in order to derive the electron density. At this point, the dissociation degree has been calculated from the emissivities $\varepsilon_{H\gamma}$, $\varepsilon_F$ respectively of the $H_\gamma$ line and the Fulcher transition:

$$n_H/n_{H2} = (\varepsilon_{H\gamma}/\varepsilon_F) \cdot [X_F(n_e,T_e)/X_{H\gamma}(n_e,T_e)]$$

where $X_{H\gamma}$ and $X_F$ are the effective emission rates of $H_\gamma$ and Fulcher transition obtained by YACORA model and by fixing $n_e$ and $T_e$ to the previous values. The calculated dissociation degree has then been used to calculate the density $n_H$ of the atomic hydrogen from the gas density $n_{H2}$, which in turn is known from the gas pressure in the source and $T_{rot}$, through the ideal gas law.

Thanks to the estimates of $n_e$ and $n_H$, $X_{H\gamma}$ has been calculated from the measured emissivity of the $H_\gamma$ line:

$$X_{H\gamma}(n_e,T_e) = \varepsilon_{H\gamma}(n_e n_H)^{-1}$$

In turn, the values of $X_{H\gamma}$ and $n_e$ have been interpreted by the CR model to obtain an updated value of $T_e$.

The entire analysis was then repeated using the updated value of $T_e$, cycling as many times as necessary to reach the convergence on the estimation of $T_e$ itself. At last, the ionization degree $\alpha$ was calculated as the ratio of $n_e$ and $n_H+n_{H2}$.

In order to characterize the NIO1 plasma source with hydrogen as filling gas, a number of power and pressure scans have been recently performed. During these scans, while raising the applied RF power, a sudden raise in the plasma luminosity was detected, of nearly one order of magnitude. These phenomena have been related to a similar source transition [9] between capacitive coupling (E-mode) and inductively coupled plasma (ICP or H-mode) for nitrogen dominated plasma. Figure 2 shows the results for the values of the main plasma parameters during an RF power scan, for a pressure of about 2.4 Pa. The gas temperature $T_{rot}$ (fig. 2a) shows a clear increase above 1400 W, from 350 °K to 550 °K. The luminosity increase (fig. 2b) happens between 1500 W and 1700 W and affects both the Fulcher band and the Balmer series emissivities. As shown in fig. 2c and 2d, at lower power the electron temperature and density are respectively of the order of some eV and $10^{17}$ $m^{-3}$; moreover the gas is weakly dissociated ($n_H/n_{H2} \approx 10^{-2}$) and even less ionized ($n_{H+}/n_H \approx 10^{-4}$), fig. 2.e. At the low values of RF power $n_e$ and

$n_{H+}/n_H$ linearly increase, while the electron temperature stays almost constant. A clear change happens at the transition to H mode at about 1400 W: the electron temperature increases from 2eV to 3 eV, the electron density and the ionization reach their top values and then start decreasing.

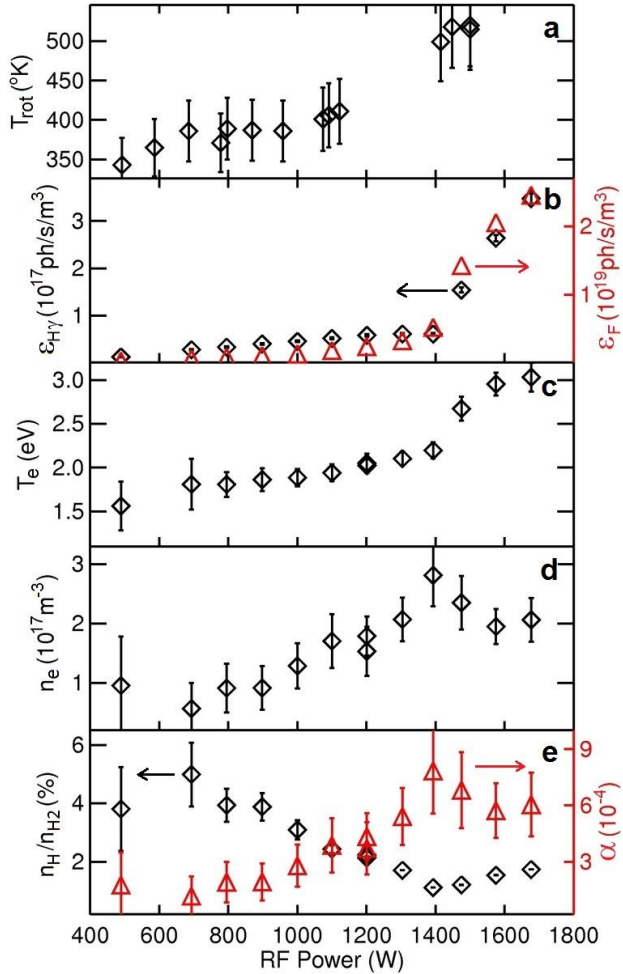

**Figure 2:** Rotational temperature $T_{rot}$ (a), $H_\gamma$ (black diamonds) and Fulcher (red triangles) emissivities (b), electron density $n_e$ (c), electron temperature $T_e$ (d), degrees of dissociation $n_H/n_{H2}$ (black diamonds) and ionization (red triangles) $\alpha$ (e) of the plasma in the NIO1 source, according to the measurements of the OES diagnostic. The data of plot (a) are referred to plasmas with gas pressures between 1.2 Pa and 2.4 Pa. The data of plots (b-e) are referred to a power scan with constant source internal pressure at 2.4 Pa.

Excluding the plasma behavior during the mode transition, the plasma dependence on applied power follows the simple general idea of an increase of the dissociation and ionization degrees as more energy is available. In this picture the decrease of $n_H/n_{H2}$ is unexpected. Since it is based on the comparison between $H_\gamma$ emission, proportional to the neutral fraction, and the Fulcher emission, proportional to the molecular fraction, the dissociation decrease is due to a stronger increase of the molecular emission than the neutral one (fig.2b). Since both emissions are similarly driven by $n_e$ and $T_e$, only a selective enhancement of the Fulcher band intensity could lead to an underestimation of $n_H/n_{H2}$. Studies are ongoing to identify the cause of this trend.

**IV. CONCLUSIONS**

An OES diagnostic was successfully installed and operated on the NIO1 negative ion source, allowing to measure the main plasma parameters at the different operational conditions set by the operator (gas pressure, RF input power, etc.). At the moment an input RF power ranging from 0.4 to 1.7 kW has been successfully coupled to a hydrogen plasma, while the filling pressure varied from 1 Pa to 7 Pa. From the high resolution spectra of the Fulcher band a rotational temperature of about 350÷500 °K has been measured. By interpreting the low resolution spectra with the population models derived from Yacora, the electron temperature, the electron density, the dissociation and ionization degrees have been estimated. The measured values of $n_e$ and $T_e$ are compatible with those detected in other negative ion sources and range respectively in the order of some eV and some $10^{17}$ m$^{-3}$, while the dissociation and the ionization degrees are lower respectively by one and two orders of magnitude [10]. The variation of the observed plasma parameters with the input RF power testifies to a plasma transition towards a full H mode, which (at a pressure of 2.4 Pa) develops in the power interval from 1400 to 1600 W. The demonstrated capability to measure the plasma parameters across the transition will allow to thoroughly investigate the underlying dynamics.

**ACKNOWLEDGMENTS**

This project has received funding from the European Union's Horizon 2020 research and innovation programme.